# An ESO–SKAO Synergistic Approach to Galaxy Formation and Evolution Studies

Isabella Prandoni[1]
Mark Sargent[2]
Elizabeth A. K. Adams[3,4]
Barbara Catinella[5,6]
Michele Cirasuolo[7]
Eric Emsellem[7,8]
Andrew Hopkins[9]
Natasha Maddox[10]
Vincenzo Mainieri[7]
Emily Wisnioski[11,6]
Matthew Colless[11]

1 INAF–Institute of Radio Astronomy, Bologna, Italy
2 International Space Science Institute, Bern, Switzerland
3 ASTRON, the Netherlands Institute for Radio Astronomy, Dwingeloo, the Netherlands
4 Kapteyn Astronomical Institute, University of Groningen, the Netherlands
5 International Centre for Radio Astronomy Research, The University of Western Australia, Crawley, Australia
6 Australian Research Council Centre of Excellence for All Sky Astrophysics in 3 Dimensions (ASTRO 3D), Australia
7 ESO
8 ENS de Lyon, CNRS, Lyon Astrophysics Research Centre, University of Lyon, France
9 School of Mathematical and Physical Sciences, Macquarie University, New South Wales, Australia
10 School of Physics, H.H. Wills Physics Laboratory, University of Bristol, UK
11 Research School of Astronomy and Astrophysics, Australian National University, Canberra, Australia

**We highlight the potential benefits of a synergistic use of SKAO and ESO facilities for galaxy evolution studies, focusing on the role that ESO spectroscopic surveys can play in supporting next-generation radio continuum and atomic hydrogen (HI) surveys. More specifically we illustrate the role that currently available or soon to be operational ESO multiplex spectrographs can play for three classes of projects: large/deep redshift survey campaigns, integral field unit/Atacama Large Millimeter/submillimeter Array (IFU/ALMA) surveys of selected regions of sky, and IFU/ALMA follow-ups of selected samples. We conclude with some general recommendations for an efficient joint exploitation of ESO–SKAO surveys.**

## Introduction

A panchromatic approach is essential for a comprehensive understanding of the complex process of galaxy formation and evolution, and of the concurrent growth of supermassive black holes (SMBH) at galaxy centres. Only through observations along the entire range of the electromagnetic spectrum it is possible to get a full census of the physical (thermal and non-thermal) processes regulating star formation and nuclear activities in galaxies, as well as of the various galaxy components (stars, multi-phase gas, dust, relativistic plasma; see, for example, Figure 1), and link these to the evolutionary properties of galaxies as a whole.

No less important is the role of multi-wavelength follow-ups of given samples to identify wavelength-dependent systematics and/or selection biases, that may hinder a full understanding of the underlying physics (see, for example, Stark et al., 2021; Catinella et al., 2023).

The astronomical community has a long track record of jointly exploiting combinations of observations at many wavelengths. Obvious examples are i) the COSMOS[1] and GOODS[2] fields, which have been targeted by virtually all telescopes from ground and space, and can count on deep photometry over a wide range of wavelengths, as well as sensitive spectroscopy, and ii) the WEAVE-LOFAR[3] project, which represents a very good example of a spectroscopic follow-up campaign of a wide-area radio-continuum survey.

In this paper we briefly review the role that ESO facilities can play in supporting galaxy-evolution studies, building on radio continuum and 21-cm line surveys from SKA and its precursors, and make some general recommendations that could render joint SKAO–ESO projects more effective (for a description of SKAO, SKA precursors and ESO facilities we refer to Bonaldi et al., on page 5 of this edition). The content of this paper is mostly (but not only) based on dedicated discussions at the Coordinated Surveys of the Southern Sky workshop[4], held in 2023 at ESO headquarters.

## SKAO and ESO working together

SKA radio continuum (Prandoni & Seymour, 2015) and 21-cm line (Staveley-Smith & Oosterloo, 2015) surveys will have a transformational impact on galaxy formation and evolution studies, as SKA precursors are already demonstrating to some extent. The former will provide an unbiased view of star formation across the Universe, and will probe jetted active galactic nuclei (AGN) down to the lowest radio powers, offering unique insights into the role of jet-induced AGN feedback. The latter will be able to detect neutral hydrogen (HI) emission from individual galaxies up to redshift $z \sim 1$ for the first time, as well as HI absorption to higher $z$, hence uncovering the role of HI in galaxy assembly and evolution.

SKAO surveys will need to be complemented by observations at shorter wavelengths to fully unlock their scientific potential, and ESO will play a key role in this respect. ESO can provide access to multi-band photometry and to optical/near-infrared (NIR) spectroscopy, through multi-object spectrographs (MOS). Both are essential for: i) properly identifying radio sources and deriving source distances (through spectroscopic or photometric redshifts); ii) classifying radio sources into AGN and star-forming galaxies (SFG); and iii) inferring important physical characteristics (for example, bolometric luminosities, stellar masses, star formation rates [SFR], metallicities, environment, etc.), and linking these to the radio-derived galaxy properties, such as, for example, SFR, AGN radio power, HI mass and kinematics, etc. Comparing radio SFR to SFR indicators derived at other wavelengths can provide useful information on selection effects due to dust attenuation. Optical/NIR spectroscopy can also play a key role in addressing galaxy/AGN HI properties beyond HI detection limits, enabling HI stacking experiments (see, for example, Brown et al., 2017; Chowdhury, Nissim & Chengalur, 2022; Sinigaglia et al., 2022).

No less important is integral field spectroscopy (IFS) and the leading role that ESO's suite of integral field units (IFU) can play in combination with Atacama Large Millimeter/submillimeter Array (ALMA) and SKAO observations. Together, they can provide a resolved view of the various components (stars, dust, molecular, ionised, atomic gas) and related processes involved in galaxy assembly and evolution. In this respect, particularly interesting is the possibility of using different tracers to probe different scales, thereby linking the internal galaxy properties with the circumgalactic medium and ultimately the cosmic web (see, for example, gas spin-filament alignments: Welker et al., 2020; Tudorache et al., 2022; Barsanti et al., 2022, 2023). Detailed investigations of the AGN fuelling/feedback cycle would also benefit greatly from resolved multi-line observations, providing 3D snapshots of inflows and outflows for the various gas phases involved in these processes. While the HI 21-cm line can be observed by the SKAO, molecular and ionised gas phases require observations at shorter wavelengths, from (sub-)mm to NIR/optical/ultraviolet bands.

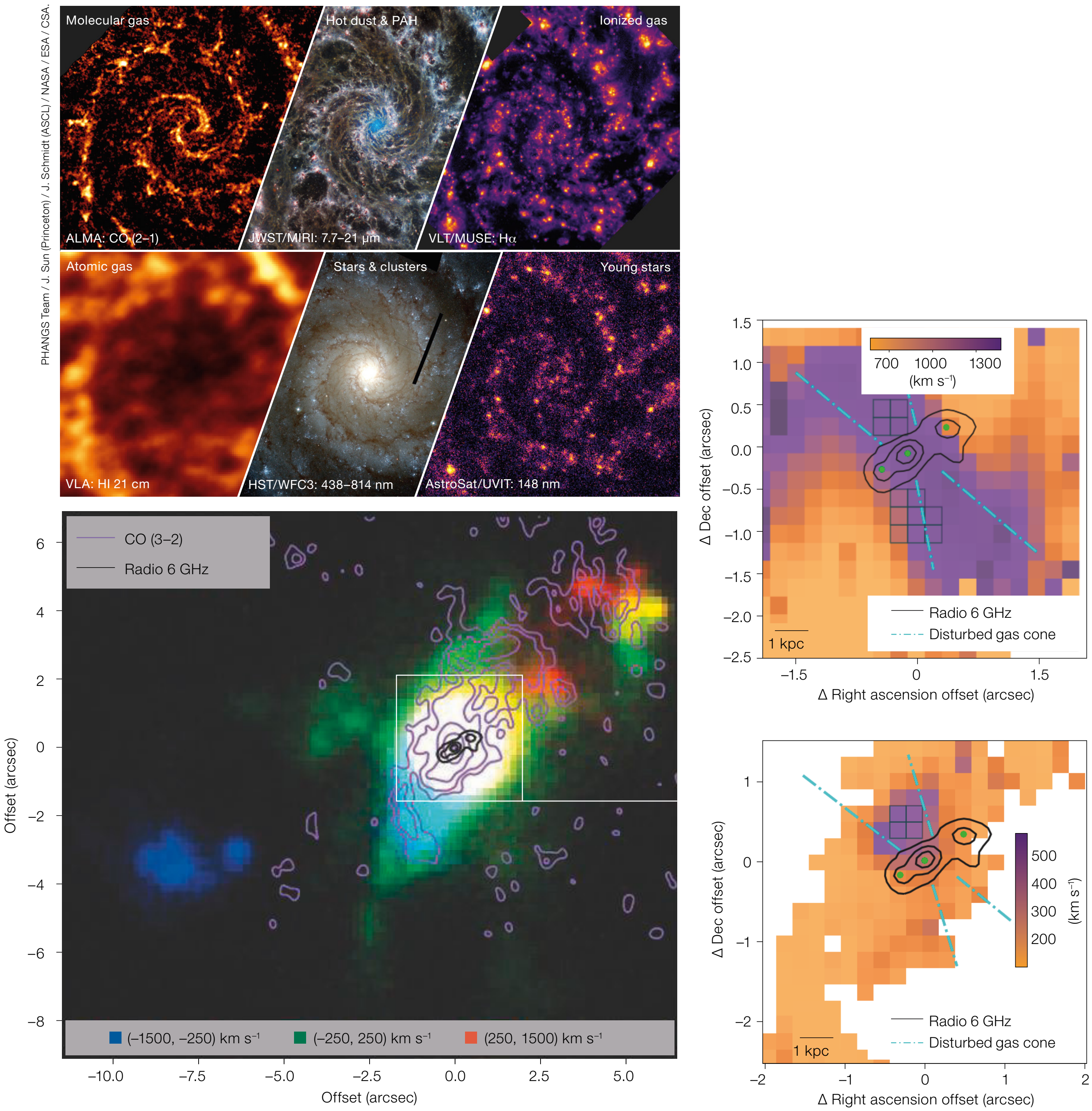

Figure 1. Top: A collection of images of the nearby disc galaxy NGC 628. Each panel highlights a different galactic constituent, obtained through observations with different instruments, over a range of wavelengths (as indicated in the panels), namely: Physics at High Angular resolution in Nearby Galaxies (PHANGS)-ALMA (Leroy et al., 2021), PHANGS-JWST (Lee et al., 2023), PHANGS-MUSE (Emsellem et al., 2022), The HI Nearby Galaxy Survey (THINGS; Walter et al., 2008), PHANGS-HST (Lee et al., 2022), and PHANGS-AstroSat (Hassani et al., 2024). Bottom left: Multi-wavelength images of a $z = 0.15$, type-2 quasar, showing the impact of small-scale jets on the interstellar medium. The background image shows three velocity slices of [OIII] λ5007 emission (ionised gas), traced by MUSE. The radio jet is shown as black contours. The right pannels show velocity dispersion of ionised gas ([OIII], top) and cold molecular gas (CO (3-2), bottom) traced by ALMA. Regions with high dispersion (purple spaxels) are orthogonal to the jet axis and different gas phases are affected on different spatial scales. Adapted from Girdhar et al. (2022).

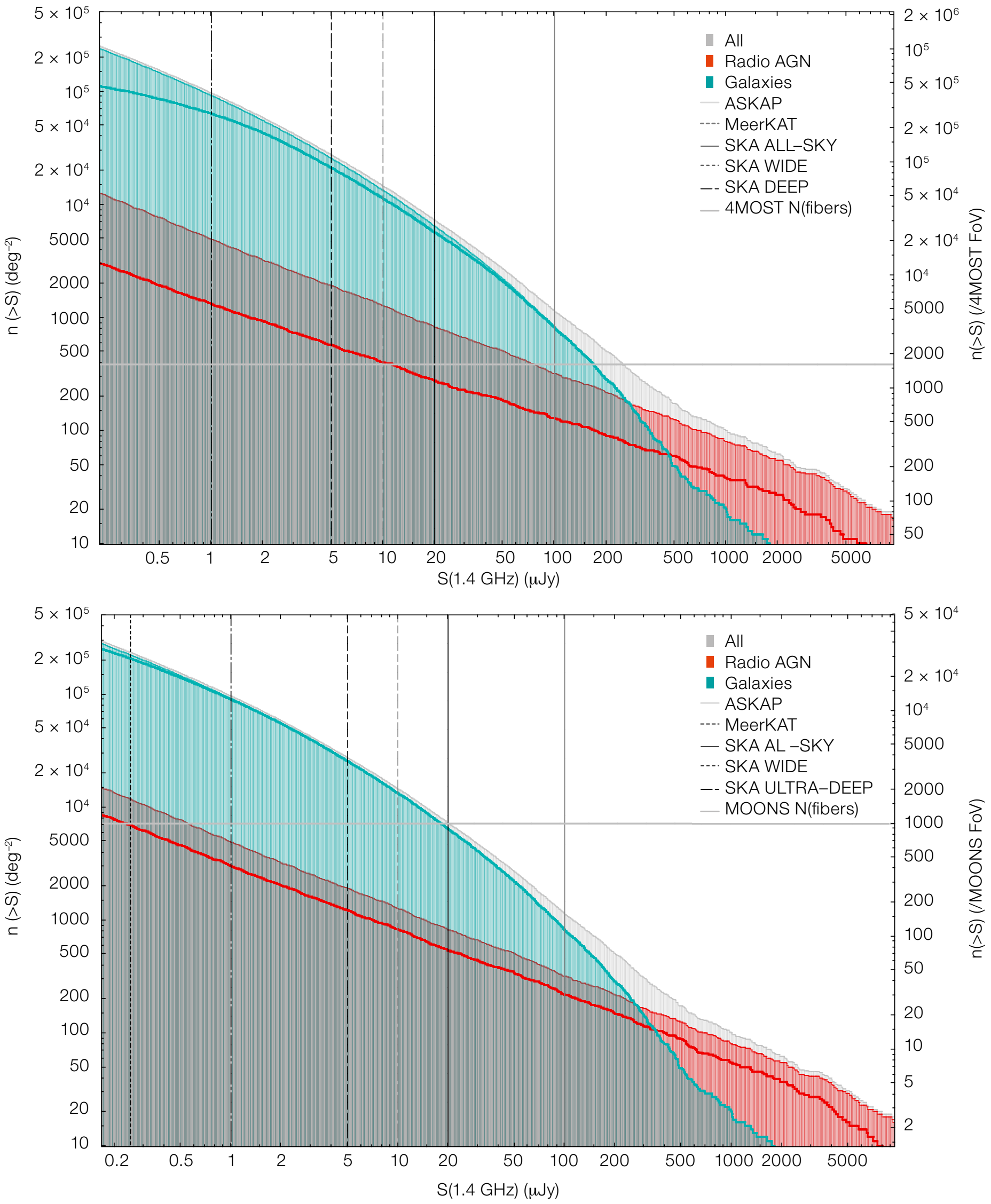

Figure 2. Radio continuum source cumulative number density predictions from T-RECS (Bonaldi et al., 2023) for different classes of sources, compared with 4MOST (top) and MOONS (bottom) capabilities. The y-axis on the left indicates the number of sources in a 1-deg$^2$ field. The y-axis on the right shows the number of sources in the 4MOST (top) and in the MOONS (bottom) field of view. The grey horizontal lines indicate the number of fibres available for 4MOST (2 × 812 for low-resolution spectroscopy; top) and MOONS (1000; bottom). The filled histograms in both panels represent radio source populations: galaxies (cyan), radio AGN (red) and the sum of the two (light grey). The thick lines of the same colours correspond to sources from the aforementioned populations that can be detected by 4MOST in two hours of exposure (i.e. with $r_{AB} < 22.5$; de Jong et al., 2019; top) or MOONS in one hour (i.e. with $H_{AB} < 22$; Cirasuolo et al., 2020; bottom). The vertical lines indicate expected radio continuum survey depths for SKA-Mid (Prandoni & Seymour, 2015) and its precursors (EMU at ASKAP and MIGHTEE at MeerKAT). In the top panel only surveys covering sky areas larger than the 4MOST field of view are shown.

Establishing joint SKAO–ESO partnerships in a timely fashion is especially important when it comes to exploiting ESO's facilities for follow-up spectroscopic campaigns, which are usually the observational bottleneck that prevents a prompt scientific exploitation of radio surveys.

We identify three main classes of spectroscopic surveys needed to support new-generation radio surveys:

- Redshift survey campaigns with medium to large field of view (FoV) multiplexed spectrographs (for example, the 4-metre Multi-Object Spectrograph Telescope [4MOST] and the Multi-Object Optical and Near-infrared Spectrograph [MOONS]), supporting wide-area and (degree-scale) deep radio continuum surveys;
- IFU (for example, the Multi Unit Spectroscopic Explorer [MUSE], BlueMUSE, the *K*-band Multi Object Spectrograph [KMOS], and the MOSAIC spectrograph at ESO's Extremely Large Telescope [ELT]) and ALMA surveys of selected regions of sky, to gather spatially resolved spectroscopy on size-able samples of galaxies and AGN for multi-phase studies on kpc/sub-kpc scales over a wide range of redshifts.
- IFU and ALMA targeted surveys (at both low and high redshift), including small FoV IFUs, such as, for example, the Enhanced Resolution Imager and Spectrograph (ERIS), the NIRSpec IFU, the Mid InfraRed Instrument's medium resolution spectroscopy mode (MIRI/MRS), and the Multi-conjugate-adaptive-optics-Assisted Visible Imager and Spectrograph (MAVIS), which trace ionised/warm molecular gas.

In addition we identify two main time scales for joint SKAO–ESO projects:

- Short- to medium-term, linked to several

upcoming opportunities such as, for example, KMOS public surveys, MOONS open time operations and second-generation 4MOST ESO community surveys (likely starting around 2030). Over this timescale the first stages of the ALMA wideband sensitivy upgrade (Carpenter et al., 2022) should also be completed;
– Long-term (> 2035), when ESO facilities will directly support SKAO surveys (and vice versa). At this point, ELT instrumentations such as MOSAIC will likely be available, perhaps further supported by a wide-field spectroscopy-dedicated telescope (like the proposed concept for the Wide-field Spectroscopic Telescope[5]; see also Mainieri et al., 2024; Bacon et al., 2024).

In the following we focus on short- to medium-term projects, that should be defined in the near future. Before the end of the current decade, spectroscopy targets will necessarily have to be selected from currently ongoing radio surveys with SKA precursors. Particularly relevant for galaxy evolution studies are ongoing surveys with SKA-Mid precursors, for example the EMU[6] and WALLABY[7] wide-area surveys at the Australian SKA Pathfinder (ASKAP), or the MIGHTEE[8] and LADUMA[9] surveys with MeerKAT.

## The role of MOS: 4MOST and MOONS

ESO's new-generation multi-fibre spectrographs 4MOST and MOONS can be effectively used for complementary spectroscopic follow-up campaigns of radio sources, targeting different redshift ranges.

Thanks to its large FoV (4.2 square degrees) and large number of fibres (1624 for low-resolution spectroscopy), 4MOST is ideally suited to follow-up spectroscopy of relatively shallow large-area radio surveys. Figure 2 shows the redshift –1.4 GHz flux density distribution of radio sources from the Tiered Radio Extragalactic Continuum Simulation (T-RECS; Bonaldi et al 2023). All radio-selected galaxies from, for example, the EMU ASKAP survey ($S_{1.4GHz}$ > 100 μJy), and from the planned all-sky SKA-Mid survey ($S_{1.4GHz}$ > 20 μJy), are detectable by 4MOST in two hours of exposure (de Jong et al., 2019; see thick cyan line in the top panel). In addition, the 4MOST fibre density (horizontal grey line) provides a good match to the cumulative number density of detectable galaxies at the EMU limit. On the other hand, in two hours 4MOST can detect only about 25–30% of the radio AGN population at all fluxes (thick red line), but provides a good match in terms of detectable radio AGN number density down to $S_{1.4GHz}$ ~ 5–10 µJy, i.e. down to the MeerKAT MIGHTEE and the planned SKA-Mid WIDE survey limits. As shown in Figure 3 (black curves), 4MOST mainly samples galaxies at $z < 1$ (top), and radio AGN at $z < 2$ (bottom). A first notable example of a synergistic MeerKAT/4MOST project is the Optical, Radio Continuum and HI Deep Spectroscopic Survey (ORCHIDSS; Duncan et al., 2023), approved for the first round of 4MOST community surveys, which pre-selected radio sources from the MeerKAT MIGHTEE, LADUMA and Fornax[10] surveys.

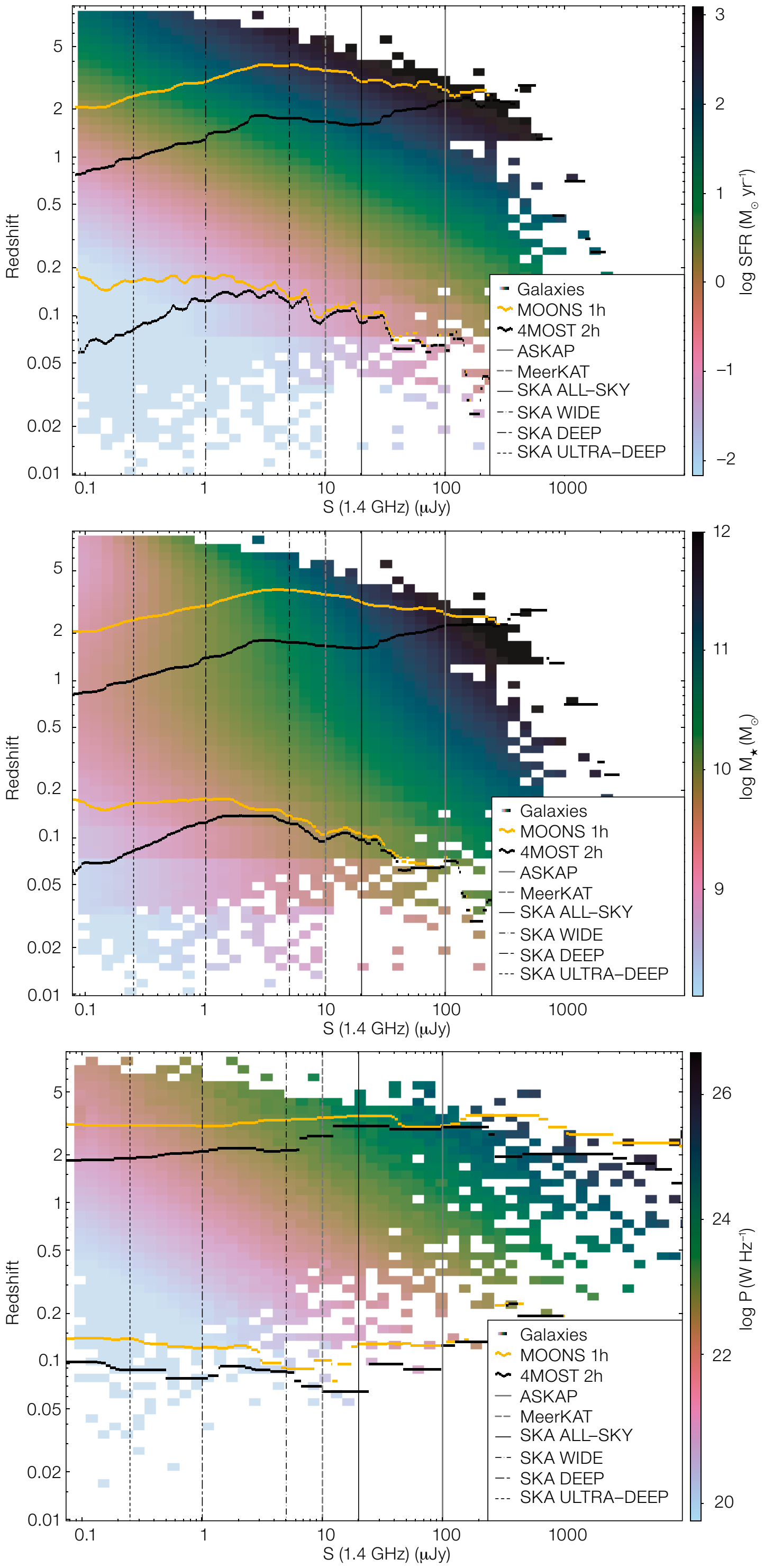

**Figure 3.** 1.4 GHz flux (in μJy) vs redshift distribution of simulated radio continuum sources from T-RECS (Bonaldi et al., 2023) for galaxies (top and middle panels) and radio-loud AGN (bottom panel) in a 1-deg$^2$ field of view. The colour grid shows variations of a third parameter, namely star formation rate and stellar mass (top and middle panels), and radio power (bottom panel). The overplotted lower and upper curves in all panels show respectively the 1% and 99% quantiles of the flux–redshift distribution for sources that can be detected by 4MOST (black curves) and MOONS (yellow curves) in the given exposure times (see legend in each panel). Vertical lines as in Figure 2.

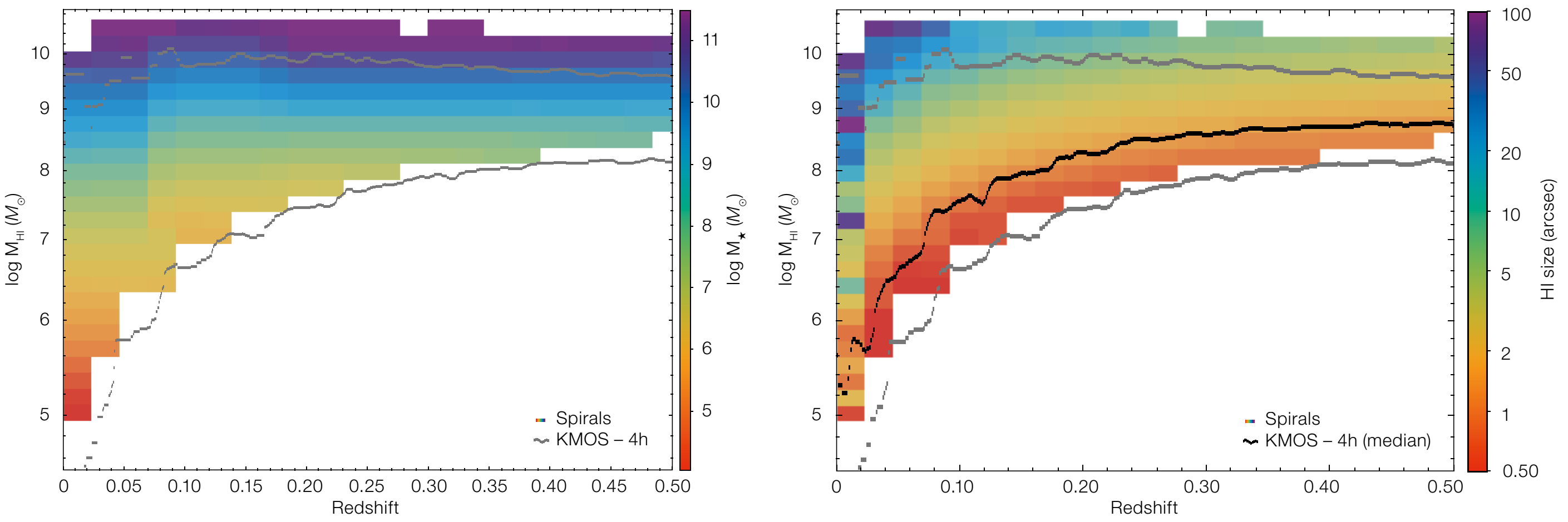

Figure 4. HI mass vs redshift distribution of late-type galaxies from T-RECS simulations (Bonaldi et al., 2023). The T-RECS HI simulation is limited to z ~ 0.5. The colour grid shows variations in stellar mass (left) and HI size (right). The overplotted grey curves show the 1% (lower) and 99% (upper) quantiles of the M(HI)–redshift distribution for sources that can be detected by KMOS with a four-hour exposure (i.e. with $K_{AB}$ < 22.5; Birkin et al., 2024). The black curve in the right panel indicates the median of the distribution.

Extending to the NIR, MOONS (1000 fibres over a 500-square-arcminute field) is particularly well suited to following up radio sources across the redshift range $1.5 < z < 2.5$, where major spectral features are redshifted out of the optical range. This redshift range is crucial for galaxy evolution studies, as it encompasses the peak of star formation and nuclear activity, and is the main target of deep, degree-scale, radio surveys with SKAO and its precursors (for example, MIGHTEE). As shown in Figure 2 (bottom), in one hour MOONS is able to detect all radio sources associated with galaxies (Cirasuolo et al., 2020; cyan line), as well as ~ 50% of the radio AGN populations (red line) down to the deepest fluxes probed by SKA-Mid radio continuum surveys ($S_{1.4GHz}$ > 0.25 µJy). The radio source cumulative number density approximately matches the MOONS fibre number density (horizontal grey line) at $S_{1.4GHz}$ ~ 10–20 µJy for galaxies and $S_{1.4GHz}$ ~ 0.2 µJy for radio AGN. It is also interesting to note that 4MOST and MOONS, combined with SKA-Mid and precursors, will be able to probe galaxies with $M_\star \sim 10^9\ M_\odot$ out to $z \sim 0.5$ and galaxies with $M_\star \sim 10^{10}\ M_\odot$ out to $z \sim 2$ (Figure 3, middle panel).

## The role of IFS and ALMA

Understanding the interplay of galaxies' physical processes requires large-area surveys and/or targeted campaigns of multi-phase gas and stellar tracers, which can be obtained through coordinated spatially resolved studies with IFUs, ALMA and SKAO. The call for KMOS public surveys[11] could be exploited to trigger joint SKAO–ESO IFS projects. With its 24 independent arms each covering a 2.8 × 2.8 arcsec$^2$ FoV, a source with major axis greater than twice the point spread function, or > 1.2 arcsec, can be considered resolved (assuming a seeing of ~ 0.6 arcsec; Birkin et al., 2024). A possible synergistic use of SKA-Mid and KMOS is illustrated in Figure 4, which shows the HI mass distribution of T-RECS simulated spirals up to redshift $z \sim 0.5$ (Bonaldi et al., 2023). All of them are detectable (Birkin et al., 2024; see grey curves), with only about half being suitable for co-spatial resolved HI-IFU studies. KMOS could thus be exploited to follow up complete HI mass-selected samples with $M(\mathrm{HI}) > 10^{7.5\text{-}8}\ M_\odot$ up to $z \sim 0.5$, with ALMA providing the coverage of the molecular gas component of the interstellar medium at similar resolutions.

The need to cover more extended sky areas and perform blind sky integrations, for specific cases like for example, deep fields, intermediate redshift galaxy clusters or the nearby star-forming galaxy population, could be supported by dedicated campaigns with monolithic wide-field IFUs such as MUSE at the Very Large Telescope (VLT) (for example, Epinat et al., 2024, Della Bruna et al., 2022). The synergistic use of blue/optical (MUSE and later BlueMUSE), sub-mm (ALMA) and radio (SKAO) facilities represents a key leverage to reveal the full complexity and multi-phase nature of the baryon cycle in galaxies.

## Summary and final considerations

As briefly discussed, ESO's MOS and IFUs can be exploited to complement new-generation galaxy-evolution-driven radio surveys. Timely radio source optical/NIR spectroscopy follow-up campaigns will ensure a prompt scientific exploitation of SKA-related surveys by, for example: i) easing the host galaxy identification and classification process; ii) enabling immediate measurements of the source physical parameters; and iii) providing prompt information on the wider environment in which the radio sources are located. IFUs and ALMA combined with SKAO will enable spatially resolved multi-scale and multiphase studies of the various processes and components involved in galaxy assembly and evolution.

In the short–medium term (≤ 2030), joint SKAO–ESO projects could exploit existing ESO instrumentation such as KMOS, MUSE and ALMA, as well as soon to be operational multi-fibre spectrographs like MOONS and 4MOST, and follow-up campaigns will have to be based on currently ongoing surveys with SKA precusors. Naturally, any planning of joint SKAO–ESO projects should bear in mind that over the next decade the optical/NIR spectroscopic survey landscape will change dramatically — not only as a consequence of 4MOST community surveys and MOONS guaranteed-time programmes, but also as a result of major non-ESO surveys such as those undertaken by ESA's Euclid mission. In the longer term, new VLT instruments like BlueMUSE, and the ELT will enable novel high-resolution and/or medium-multiplex spectroscopic studies. For specific radio source populations (for example, optically faint radio-loud objects) the ELT with MOSAIC can reveal the presence of outflow components and characteristics, and targeted IFU observations with the High Angular Resolution Monolithic Optical and Near-infrared Integral field spectrograph (HARMONI) will permit fine-grained studies of the astrophysics of stellar and AGN feedback processes within galaxies. A future facility like the WST has the potential to provide optical counterparts for the SKAO over large (~ 3 $deg^2$) MOS survey areas, and to study the gaseous and stellar emission in the environment of, for example, AGN at cosmic noon ($z$ ~ 1–3) thanks to its IFU, opening new scientific opportunities and synergies.

As a final remark, we note that a prompt scientific exploitation of joint ESO–SKAO surveys will likely require an efficient use of the available infrastructures. From an operational point of view, we therefore encourage SKAO–ESO coordinated efforts towards joint proposal schemes, shared archival capabilities and joint virtual observing platforms, possibly building on the work already done by ESO and SKAO/SKA-precursors teams.

## Acknowledgements

We thank all the collegues who attended the splinter session Galaxies and Galaxy Evolution of the 2023 Coordinated Surveys of the Southern Sky workshop, for their active participation. This paper is based heavily on the outcomes of the splinter discussion sessions.

## Links

[1] COSMOS: https://cosmos.astro.caltech.edu
[2] GOODS: https://www.stsci.edu/science/goods/
[3] WEAVE-LOFAR: https://arxiv.org/abs/1611.02706
[4] Coordinated Surveys of the Southern Sky workshop: https://www.eso.org/sci/meetings/2023/CSSS.html
[5] WST: https://www.wstelescope.com/
[6] EMU: http://emu-survey.org/
[7] WALLABY: https://wallaby-survey.org
[8] MIGHTEE: https://www.mighteesurvey.org/
[9] LADUMA: https://science.uct.ac.za/laduma
[10] MeerKAT Fornax Survey: https://sites.google.com/inaf.it/meerkatfornaxsurvey
[11] KMOS public survey call: https://www.eso.org/sci/observing/PublicSurveys/KMOSloicall.html

SKAO

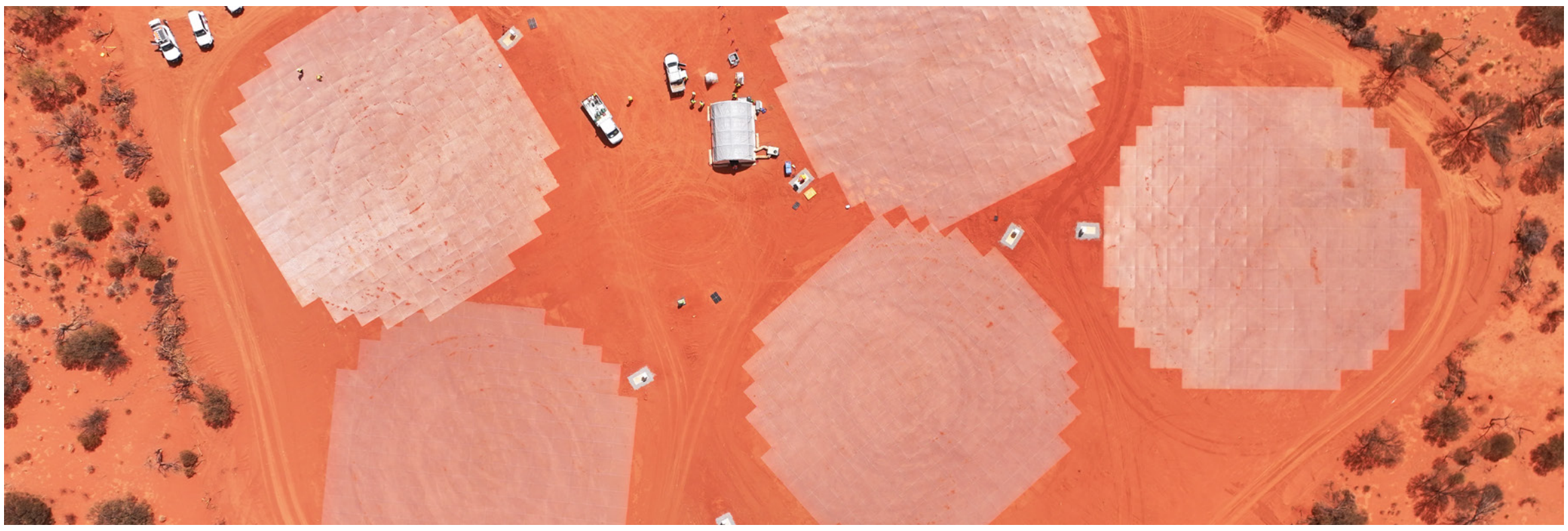

Bird's eye view of S8, a cluster of SKA-Low antenna stations. Image taken in March 2024.